\documentclass[twocolumn,prl,aps,showpacs,superscriptaddress,floatfix,showkeys,nofootinbib,10pt]{revtex4-1}

\usepackage[T1]{fontenc}

\usepackage{amsmath,amsfonts,amssymb,amsthm}
\usepackage{graphicx}
\usepackage{bbm,bm}

\newcommand{\be}{\begin{equation}} \newcommand{\ee}{\end{equation}}
\newcommand{\ba}{\begin{aligned}} \newcommand{\ea}{\end{aligned}}

\DeclareMathOperator{\Tr}{Tr}

\begin{document}

\title{Experimentally feasible semi-device-independent certification of $4$ outcome POVMs}

\author{Piotr~Mironowicz} \email{piotr.mironowicz@gmail.com}
\affiliation{Department of Algorithms and System Modeling, Faculty of Electronics, Telecommunications and Informatics, Gda\'nsk University of Technology}
\affiliation{Institute of Theoretical Physics and Astrophysics, National Quantum Information Centre, Faculty of Mathematics, Physics and Informatics, University of Gda\'nsk, Wita Stwosza 57, 80-308 Gda\'nsk, Poland}

\author{Marcin~Paw{\l}owski}
\affiliation{Institute of Theoretical Physics and Astrophysics, National Quantum Information Centre, Faculty of Mathematics, Physics and Informatics, University of Gda\'nsk, Wita Stwosza 57, 80-308 Gda\'nsk, Poland}
\affiliation{Instytut Fizyki Teoretycznej i Astrofizyki, Uniwersytet Gda\'{n}ski, PL-80-952 Gda\'{n}sk, Poland}

\date{\today}

\begin{abstract}
In this paper we construct a semi-device-independent protocol able to certify the presence of the generalized measurements. We show robustness of the protocol and conclude that it allows for experimental realisations using current technology.
\end{abstract}


\maketitle

\textit{Introduction.-}
Measurements lay in the heart of all physical sciences. The formalism of quantum mechanics defines them as projections on vectors in some abstract Hilbert space, yet they remain one of the most mysterious notions of the modern physics.

Even though the formalism was elaborated many years ago~\cite{Dirac}, the development of quantum information science lead to its significant enhancement, when so-called positive-operator valued measure (POVM) or generalised measurements have been introduced~\cite{Helstrom,Ivanovic,Peres}, originating in the problem of non-orthogonal states discrimination.

A recent trend in quantum information, related to security and reliability issues, tries to formulate physical problems, protocols and experiments in so-called device-independent (DI) way~\cite{DI}, where one derives all conclusions about some phenomenon basing only on the observed results, not making assumptions regarding internal construction of devices involved. A popular relaxation of this approach is called semi-device-independent (SDI)~\cite{SDI}, where some assumptions regarding devices are made. The most common assumption is a constraint on the dimension of quantum systems communicated between parts of the considered setup.

From these two topic emerged a problem of SDI certification that some of the measurements performed by devices used in an experiment are POVMs. Surprisingly, the problem revealed to be very demanding. The first experiment certifying presence of generalized measurements~\cite{DI-qubit} was based on a scheme involving entangled quantum states. Later similar approach was used to generate randomness from POVM outcomes~\cite{DI-rand}. Yet, there was not known a method how to certify presence of generalized measurements without entangled states.

The result of this paper is an SDI protocol using the prepare and measure scheme that is able to certify presence of a $4$-outcome POVM in dimension two. We show that the robustness of the protocol makes it feasible for experimental realisations.

\textit{Quantum Random Access Codes.-}
A key tool we use in this paper are so-called \textit{Random Access Codes} (RACs)~\cite{RACs}. In an $n^d \rightarrow 1$ RAC Alice is provided a uniformly distributed string of $n$ dits, $\mathbf{x} = x_1 x_2 \cdots x_{n}$. Her task is to encode the string into a single dit message $m(\mathbf{x})$, in such a way that Bob is able to recover any of the $n$ dits with high probability. Bob receives Alice's message $m$, and a referee provides him a uniformly distributed input $y\in [n]$, where $[n] \equiv \{1,2,\dots,n\}$. After this Bob performs classical (possibly probabilistic) computations of some function $b(y, m)$. We say that the protocol succeeded when $b(y, m(\mathbf{x})) = x_y$.

The quantum analog of RACs are Quantum Random Access Codes (QRACs)~\cite{RACs}. In $n^d \rightarrow 1$ QRAC Alice encodes the $n$-dit input $\mathbf{x}$ into a $d$-dimensional quantum system $\rho_{\mathbf{x}}$, that is then transmitted to Bob. He afterwards performs one of his $n$ measurements (depending on the input $y$) to give his guess $b$ for $x_y$. Thus he guesses $b$ with probability given by $\Tr[\rho_{\mathbf{x}} M^y_b]$, where the operators $M^y_b$ are POVMs, i.e. positive and $\forall y \sum_b M^y_b = \openone_d$, and $\openone_d$ is the identity operator in dimension $d$. In this paper we consider degenerated POVMs, where some of the operators are allowed to be null.

Note that Alice's only optimal strategy is to send a state $\rho_{\mathbf{x}}$ maximizing the value of $\Tr \left[ \rho_{\mathbf{x}} \left( \sum_{y \in [n]} M^{y}_{x_y} \right) \right]$. This is obtained if the state is in the subspace of vectors with maximal eigenvalue of the operator $\sum_{y \in [n]} M^{y}_{x_y}$.

In both RACs and QRACs we consider the probability that Bob returns outcome $b$ when his input is $y$, and Alice's input is $x$. We denote this probability by $P(b|x,y)$. The average success probability for RACs and QRACs is thus given by:
\be
	\label{eq:psucc}
	\frac{1}{nd^n} \sum_{\mathbf{x} \in [d]^n} \sum_{y \in [n]} P(x_y|\mathbf{x}, y).
\ee

\textit{Reduction of $3^2 \rightarrow 1$ QRAC and its self-testing.-}
Now, let us focus on a $3^2 \rightarrow 1$ QRAC. In~\cite{Armin18} it has been shown that this protocol posses a robust self-testing property~\cite{selftesting}, meaning that there exist a unique set of optimal quantum states and measurements that maximizes it (up to some elementary transformations), and that if the experiment is close to the maximal value, then the states and measurement are close to the optimal ones. Let
\be
	\label{eq:optimal}
	\{\tilde{\rho}_\mathbf{x}\} \text{ and } \{\tilde{M}^y_b\}
\ee
be these optimal states and measurements, respectively.

To be more specific, Alice obtains here one of $8$ possible inputs, and prepares one of $8$ relevant qubits. In the perfect case of success probability
\be
	\label{eq:succ3to1Value}
	S_3 = \frac{1}{2} \left(1 + \frac{\sqrt{3}}{3}\right) \approx 0.78868
\ee
Bob performs a measurement in one of $3$ MUBs in dimension $2$, corresponding in this case to measuring observables given by Pauli operators $\{\sigma_x, \sigma_y, \sigma_z\}$.

From our observation regarding Alice's optimal encodings we find the unique preparation states maximizing the success probability. One can check~\cite{RACs} that the Bloch sphere representations of states $\{\tilde{\rho}_{000}, \tilde{\rho}_{011}, \tilde{\rho}_{101}, \tilde{\rho}_{110}\}$ and $\{\tilde{\rho}_{111}, \tilde{\rho}_{100}, \tilde{\rho}_{010}, \tilde{\rho}_{001}\}$ are located on the edges of regular tetrahedra, with relevant edges antipodal to each other. We provide the explicit representation of the former set of states in the Appendix in~\eqref{eqs:states}.

Now, we apply the method of the so-called reduction of symmetric dimension witnesses introduced by us in~\cite{dimwits} to show that the following expression (that is not a QRAC) posses a self-testing property:
\be
	\label{eq:reduced}
	\frac{1}{12} \sum_{\mathbf{x} \in \{000,011,101,110\}} \sum_{y \in [3]} P(x_y|\mathbf{x}, y).
\ee
Indeed, let us assume that there exist a set of measurements of Bob $\{M^y_b\}$ and states prepared by Alice $\{\rho_\mathbf{x}\}_{\mathbf{x} = 000,011,101,110}$ optimal for~\eqref{eq:reduced} and different from $\{\tilde{M}^y_b\}$ and $\{\tilde{\rho}_\mathbf{x}\}_{\mathbf{x} = 000,011,101,110}$. Without loss of generality we may assume that $\Tr M^y_b = 1$ for all $b$ and $y$~\cite{Armin18}.

Let us now define $\rho_{111} \equiv \openone_2 - \rho_{000}$, $\rho_{100} \equiv \openone_2 - \rho_{011}$, $\rho_{010} \equiv \openone_2 - \rho_{101}$ and $\rho_{001} \equiv \openone_2 - \rho_{110}$. One can easily check that these states together with $\{M^y_b\}$ maximizes the expression complementary to~\eqref{eq:reduced}, i.e.
\be
	\label{eq:reducedComplementary}
	\frac{1}{12} \sum_{\mathbf{x} \in \{111,100,010,001\}} \sum_{y \in [3]} P(x_y|\mathbf{x}, y).
\ee

One can verify that the states $\{\tilde{\rho}_\mathbf{x}\}$ and measurements $\{\tilde{M}^y_b\}$ from~\eqref{eq:optimal} used for expressions~\eqref{eq:reduced} and~\eqref{eq:reducedComplementary} give for both cases success probability $S_3$, and the average of these game expressions is exactly $3^2 \rightarrow 1$ QRAC. From its self-testing property, the assumption that states $\{\rho_\mathbf{x}\}$ and measurements $\{M^y_b\}$ are not equal to these leads to contradiction, showing the self-testing property of both games~\eqref{eq:reduced} and~\eqref{eq:reducedComplementary}.

We briefly note here that an immediate consequence of the above construction is the possibility of deriving also the robustness of the self-testing of expressions~\eqref{eq:reduced} and~\eqref{eq:reducedComplementary} directly from robustness of the $3^2 \rightarrow 1$ QRAC~\cite{Armin18}. Indeed, consider the maximal distance $\delta_1$ of states and measurements from~\eqref{eq:optimal} that allows to reach the value $S_3-\epsilon$ by $3^2 \rightarrow 1$ QRAC, where to express the distance an arbitrary isotropic metric is used.

Let $\delta_2$ be the maximal distance from~\eqref{eq:optimal} that allows to reach the value $S_3-\epsilon$ in the reduced game~\eqref{eq:reduced}. From isotropy of the metric we get the same maximal distance $\delta_2$ for the reduced game~\eqref{eq:reducedComplementary}, and we see that the same measurements can be used for both of these reduced games. From this and from the fact that $3^2 \rightarrow 1$ QRAC is the average of~\eqref{eq:reduced} and~\eqref{eq:reducedComplementary} we see that $\delta_2 \leq \delta_1$.

\textit{Robust POVM certification.-}
Let us now consider a more complicated task, where Alice and Bob are not only maximizing the expression~\eqref{eq:reduced} (i.e. the reduced $3^2 \rightarrow 1$ QRAC), but also minimizing probability of some other events. Let us introduce an additional $4$th input of Bob, related to a $4$-outcome measurement (with outcomes labelled $1,2,3,4$). The new expression we consider is:
\be
	\label{eq:certificate}
	\frac{1}{12} \left[ \sum_{\mathbf{x} \in \{000,011,101,110\}} \sum_{y \in [3]} P(x_y|\mathbf{x}, y) - k G^4\right],
\ee
where $k > 0$, and
\be
	\label{eq:G4}
	G^4 \equiv P(1|000,4) + P(2|011,4) + P(3|101,4) + P(4|110,4).
\ee
One can easily note, that the expression~\eqref{eq:certificate} cannot obtain value greater than $S_3$, and the value would be obtained only when the states $\{\tilde{\rho}_\mathbf{x}\}$ and measurements $\{\tilde{M}^y_b\}$ are used, and the part $G^4$ is equal $0$. From this it follows that each operator $\{M^4_i\}$ has to be orthogonal to relevant state $\{\tilde{\rho}_{000}, \tilde{\rho}_{011}, \tilde{\rho}_{101}, \tilde{\rho}_{110}\}$. Direct calculation shows that the $4$th measurement is a POVM specified by operators given in the Appendix in~\eqref{eqs:POVM}.

All the above considerations referred to the perfect case when the maximal value $S_3$ of the expression~\eqref{eq:certificate} is observed. In real world experiment this will not be the case due to experimental imperfections. Thus, to provide an experimentally feasible certificate that a measurement is a $4$-outcome POVM, we need to establish the robustness of the certification protocol. We note here, that in order to perform a conclusive experiment one does not need to calculate full robustness properties including distances of the states and measurements to the perfect one depending on the certificate value. For the purpose of the experiment it suffices to establish the scope of values of the certificate that guaranties the presence of $4$ or $3$ outcome POVM.

In order to model the experimental imperfections we use the visibility of quantum states. The visibility $\nu$ means that whenever the state prepared in the perfect situation is $\rho$, the state occurring in the experiment is $\nu \rho + \frac{1-\nu}{d} \openone_d$. This parameter represents the impact of the proper state in comparison to the white noise. Let $N(k)$ be the value of the certificate~\eqref{eq:certificate} when all states are completely noised. We have
\be
	N(k) = \frac{1}{12} (12 \cdot 0.5 - k \cdot 4 \cdot 0.25).
\ee
Let $g_j(k)$ denote the maximal value of the expression~\eqref{eq:certificate} when the $4$th measurement has $j$ outcomes, $j = 2, 3, 4$. We have $g_4(k) = S_3$ for all $k \geq 0$. The critical visibility $\nu_j(k)$ needed to certify that $j$, $j=3, 4$, outcomes are necessary to reproduce the experimental value are given by the expression:
\be
	\nu_j(k) = \frac{g_{j-1}(k) - N}{S_3 - N}.
\ee

There exist a couple of methods of optimization of quantum expression with dimension constraints. One of them is the see-saw method~\cite{seesaw}, that optimizes within interior of the quantum theory, and is able to provide explicit form of states and measurements realizing the resulting value. The result of a maximization provides a lower-bound on the proper quantum value. Other methods optimize from exterior of the quantum theory (i.e. they provide relaxations of the quantum formalism) and include \cite{dimwits,NTV,MV}. These methods provide upper-bounds on results of maximizations. Recall that the assumed dimension constraint is two.

\begin{figure}[htbp!]
	\includegraphics[width=\linewidth]{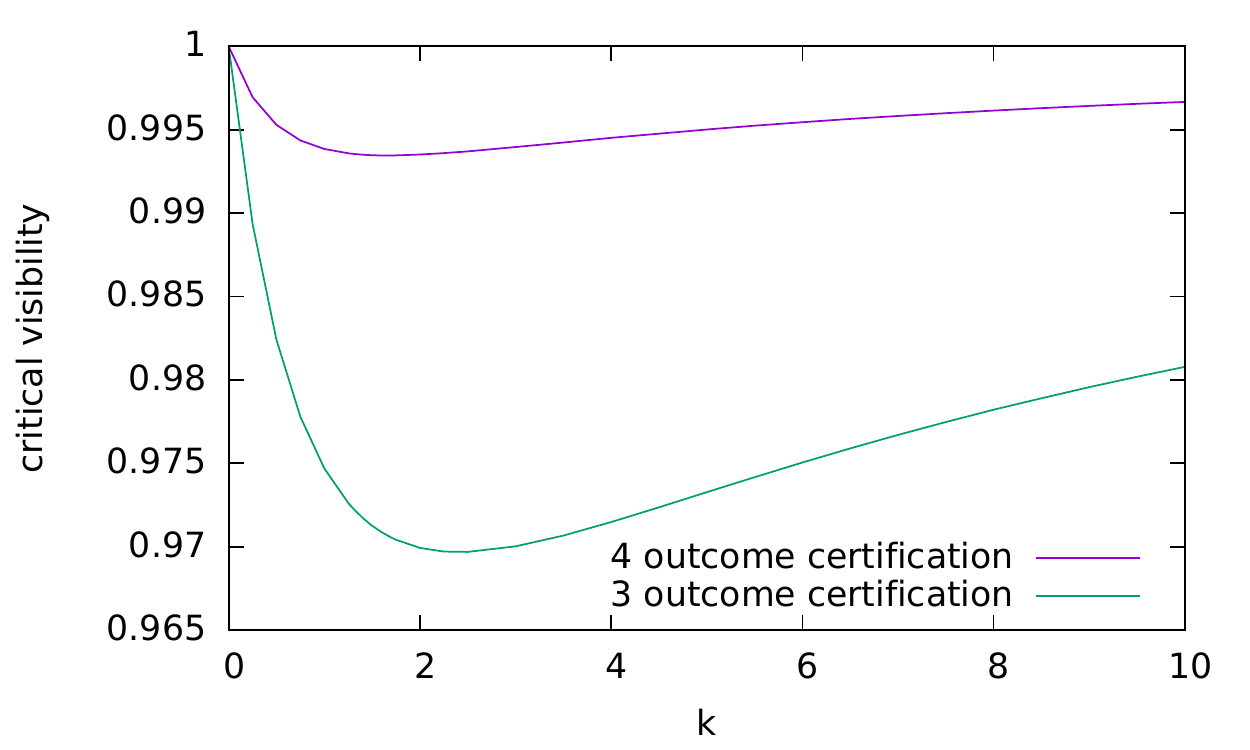}
	\caption{The values of critical visibilities needed to certify that a general measurement has been used in an experiment involving the game~\eqref{eq:certificate} for different values of $k$ parameter.}
	\label{fig:visibilities}
\end{figure}


The results we obtained using the second level of the MLP relaxation~\cite{dimwits} for $k = 1$ are $g_3 = 0.7864$ and $g_2 = 0.7793$, relating to visibilities $\nu_4 = 0.9938$ and $\nu_3 = 0.9747$, respectively. The value $g_2$ is derived also in the see-saw method giving lower-bounds on the values of dimension constrained quantum scenarios, thus this value is exact. In the see-saw approximation of $g_3$ we obtained after multiple executions with random seeds the value $0.7856$, thus establishing the scope of the possible exact value.

Fig.~\ref{fig:visibilities} shows critical visibilities obtained using the MLP relaxation for different values of $k$. The above result means that whenever in an experiment one obtains the average value grater than $g_3$ ($g_2$), then a $4$($3$)-outcome POVM is certified to be used. The visibilities $\nu_4$ and $\nu_3$ seem to be experimentally feasible. For the sake of completeness the details of the probability distribution obtained using see-saw are given in the Appendix in Tabs~\ref{tab:2outcomes} and~\ref{tab:3outcomes} for $k = 1$.

\textit{Conclusions.-}
In this paper we have presented a prepare and measure SDI protocol able to certify the presence of $4$ outcome generalized measurements in dimension two. The robustness of the protocol allows for using it in real world experiments in laboratories~\cite{ArminPnM}.

Even though the construction of the protocol was based on reduction of QRACs, the resulting states and measurements are closely related to so-called Elegant Bell Inequality~\cite{EBI} (EBI), which self-testing properties were shown recently~\cite{selftestingEBI}. Using the methods of~\cite{dimwits} it is possible to convert EBI to a prepare and measure protocol and, after the reduction operation, derive the same result as those presented above.

The problem how to certify generalized measurements in different dimensions and with arbitrary number of outcomes remains open. The above construction suggests that a possible way to tackle this issue is related to similar QRAC constructions~\cite{myPreparation}. This shows the benefit of using the above method in comparison to a simple conversion from existing entangled protocol using EBI.

\textit{Acknowledgements.-}
During the completion of this manuscript, we became aware of an independent work~\cite{ArminPnM} focused on selftesting of qubit POVMs.

The paper was supported by National Science Centre (NCN) Grant No. 2014/14/E/ST2/00020, First TEAM (Grant No. First TEAM/2016-1/5) and DS Programs of the Faculty of Electronics, Telecommunications and Informatics, Gda\'nsk University of Technology. The see-saw optimizations have been performed using OCTAVE \cite{octave} with SDPT3 solver \cite{SDPT3a,SDPT3b} and YALMIP toolbox \cite{yalmip}. The MLP relaxation values have been calculated using Ncpol2spda package~\cite{ncpol2sdpa}.

\begin{appendix}

Let $\alpha \equiv \frac{3 - \sqrt{3}}{6}$ and $\beta \equiv \frac{\sqrt{3}}{6}$.

The state preparation are given by:
\begin{subequations}
	\label{eqs:states}
	\be
	\tilde{\rho}_{000} =
	\begin{bmatrix}
		1 - \alpha & \beta (1 + i) \\
		\beta (1 - i) & \alpha
	\end{bmatrix}
	\ee
	\be
	\tilde{\rho}_{011} =
	\begin{bmatrix}
		\alpha & \beta (1 - i) \\
		\beta (1 + i) & 1 - \alpha
	\end{bmatrix}
	\ee
	\be
	\tilde{\rho}_{101} =
	\begin{bmatrix}
		1 - \alpha & \beta (-1 + i) \\
		\beta (-1 - i) & \alpha
	\end{bmatrix}
	\ee
	\be
	\tilde{\rho}_{110} =
	\begin{bmatrix}
		1 - \alpha & \beta (-1 - i) \\
		\beta (-1 + i) & \alpha
	\end{bmatrix}
	\ee
\end{subequations}

The 4 outcome POVM used as the $4$th measurement in the proposed protocol is given by
\begin{subequations}
	\label{eqs:POVM}
	\be
 	M^4_1 = 
 		\frac{1}{2}
 		\begin{bmatrix}
 			\alpha & \beta (-1-i) \\
 			\beta (-1+i) & 1-\alpha
 		\end{bmatrix}
 	\ee
 	\be
 	M^4_2 = 
 		\frac{1}{2}
	 	\begin{bmatrix}
		 	1-\alpha & \beta (-1+i) \\
		 	\beta (-1-i) & \alpha
	 	\end{bmatrix}
 	\ee
 	\be
	M^4_3 = 
		\frac{1}{2}
		\begin{bmatrix}
			1-\alpha & \beta (1-i) \\
			\beta (1+i) & \alpha
		\end{bmatrix}
	\ee
	\be
	M^4_4 = 
		\frac{1}{2}
		\begin{bmatrix}
			\alpha & \beta (1+i) \\
			\beta (1-i) & 1-\alpha
		\end{bmatrix}
	\ee
\end{subequations}

\begin{widetext}
	\begin{table}[htbp!]
		\caption{The probabilities obtained using see-saw method for the case when $4$th measurement is assumed to have only \textit{two outcomes} with non-zero probabilities and $k = 1$. The bold values are the ones that occur in~\eqref{eq:certificate}.}
		\label{tab:2outcomes}
		\begin{tabular}{|r||c|c||c|c||c|c||c|c|c|c|}
			\hline
			$\mathbf{x}$ & $P(0|\mathbf{x},1)$ & $P(1|\mathbf{x},1)$ & $P(0|\mathbf{x},2)$ & $P(1|\mathbf{x},2)$ & $P(0|\mathbf{x},3)$ & $P(1|\mathbf{x},3)$ & $P(1|\mathbf{x},4)$ & $P(2|\mathbf{x},4)$ & $P(3|\mathbf{x},4)$ & $P(4|\mathbf{x},4)$ \\
			\hline
			$000$ & $\mathbf{0.684}$ & $0.316$ & $\mathbf{0.854}$ & $0.146$ & $\mathbf{0.854}$ & $0.146$ & $\mathbf{0.035}$ & $0.965$ & $0$ & $0$ \\
			\hline
			$011$ & $\mathbf{0.684}$ & $0.316$ & $0.146$ & $\mathbf{0.854}$ & $0.146$ & $\mathbf{0.854}$ & $0.965$ & $\mathbf{0.035}$ & $0$ & $0$ \\
			\hline
			$101$ & $0.195$ & $\mathbf{0.805}$ & $\mathbf{0.757}$ & $0.243$ & $0.243$ & $\mathbf{0.757}$ & $0.500$ & $0.500$ & $\mathbf{0}$ & $0$ \\
			\hline
			$110$ & $0.195$ & $\mathbf{0.805}$ & $0.243$ & $\mathbf{0.757}$ & $\mathbf{0.757}$ & $0.243$ & $0.500$ & $0.500$ & $0$ & $\mathbf{0}$ \\
			\hline
		\end{tabular}
	\end{table}
	\begin{table}[htbp!]
		\caption{The probabilities obtained using see-saw method for the case when $4$th measurement is assumed to have \textit{three outcomes} with non-zero probabilities and $k = 1$. The bold values are the ones that occur in~\eqref{eq:certificate}.}
		\label{tab:3outcomes}
		\begin{tabular}{|r||c|c||c|c||c|c||c|c|c|c|}
			\hline
			$\mathbf{x}$ & $P(0|\mathbf{x},1)$ & $P(1|\mathbf{x},1)$ & $P(0|\mathbf{x},2)$ & $P(1|\mathbf{x},2)$ & $P(0|\mathbf{x},3)$ & $P(1|\mathbf{x},3)$ & $P(1|\mathbf{x},4)$ & $P(2|\mathbf{x},4)$ & $P(3|\mathbf{x},4)$ & $P(4|\mathbf{x},4)$ \\
			\hline
			$000$ & $\mathbf{0.765}$ & $0.235$ & $\mathbf{0.765}$ & $0.235$ & $\mathbf{0.856}$ & $0.144$ & $\mathbf{0.008}$ & $0.496$ & $0.496$ & $0$ \\
			\hline
			$011$ & $\mathbf{0.765}$ & $0.235$ & $0.144$ & $\mathbf{0.856}$ & $0.235$ & $\mathbf{0.765}$ & $0.496$ & $\mathbf{0.008}$ & $0.496$ & $0$ \\
			\hline
			$101$ & $0.144$ & $\mathbf{0.856}$ & $\mathbf{0.765}$ & $0.235$ & $0.235$ & $\mathbf{0.765}$ & $0.496$ & $0.496$ & $\mathbf{0.008}$ & $0$ \\
			\hline
			$110$ & $0.235$ & $\mathbf{0.765}$ & $0.235$ & $\mathbf{0.765}$ & $\mathbf{0.765}$ & $0.235$ & $0.333$ & $0.333$ & $0.333$ & $\mathbf{0}$ \\
			\hline
		\end{tabular}
	\end{table}
\end{widetext}

\end{appendix}

\end{document}